\documentstyle[a4,12pt]{article}

\newcommand{\nc}[2]{\newcommand{#1}{#2}}
\newcommand{\ncx}[3]{\newcommand{#1}[#2]{#3}}
\ncx{\pr}{1}{#1^{\prime}}
\nc{\nl}{\newline}
\nc{\np}{\newpage}
\nc{\nit}{\noindent}
\nc{\be}{\begin{equation}}
\nc{\ee}{\end{equation}}
\nc{\ba}{\begin{array}}
\nc{\ea}{\end{array}}
\nc{\dsp}{\displaystyle}
\nc{\bit}{\bibitem}
\nc{\ct}{\cite}
\ncx{\dd}{2}{\frac{\partial #1}{\partial #2}}
\nc{\pl}{\partial}
\nc{\dg}{\dagger}
\nc{\ag}{\alpha}
\nc{\bg}{\beta}
\nc{\gam}{\gamma}
\nc{\Gam}{\Gamma}
\nc{\bgm}{\bar{\gam}}
\nc{\del}{\delta}
\nc{\Del}{\Delta}
\nc{\eps}{\epsilon}
\nc{\ve}{\varepsilon}
\nc{\th}{\theta}
\nc{\vt}{\vartheta}
\nc{\kg}{\kappa}
\nc{\lb}{\lambda}
\nc{\ps}{\psi}
\nc{\Ps}{\Psi}
\nc{\sg}{\sigma}
\nc{\spr}{\pr{\sg}}
\nc{\Sg}{\Sigma}
\nc{\rg}{\rho}
\nc{\fg}{\phi}
\nc{\Fg}{\Phi}
\nc{\vf}{\varphi}
\nc{\og}{\omega}
\nc{\Og}{\Omega}
\nc{\Kq}{\mbox{$K(\vec{q},t|\pr{\vec{q}\,},\pr{t})$ }}
\nc{\Kp}{\mbox{$K(\vec{q},t|\pr{\vec{p}\,},\pr{t})$ }}
\nc{\vq}{\mbox{$\vec{q}$}}
\nc{\qp}{\mbox{$\pr{\vec{q}\,}$}}
\nc{\vp}{\mbox{$\vec{p}$}}
\nc{\va}{\mbox{$\vec{a}$}}
\nc{\vb}{\mbox{$\vec{b}$}}
\nc{\Ztwo}{\mbox{\sf Z}_{2}}
\nc{\Tr}{\mbox{Tr}}
\nc{\lh}{\left(}
\nc{\rh}{\right)}
\nc{\ld}{\left.}
\nc{\rd}{\right.}
\nc{\cB}{\mbox{$^{\ast}\Og$ }}
\nc{\nil}{\emptyset}
\nc{\bor}{\overline}
\nc{\cD}{{\cal D}}
\nc{\tQ}{\tilde{Q}}
\nc{\wig}{\wedge}

\begin{document}

\begin{flushright}
NIKHEF-H/94-28
\end{flushright}
\vspace{3ex}

\begin{center}

{\LARGE {\bf Killing-Yano tensors and generalized}} \\
\vspace{3ex}

{\LARGE {\bf  supersymmetries in black-hole and}} \\
\vspace{3ex}

{\LARGE {\bf monopole geometries\footnote{Contribution to the proceedings
             of Quarks-94; Vladimir, Russia (1994)}  }} \\
\vspace{5ex}

{\Large J.W. van Holten } \\
\vspace{3ex}

{\Large NIKHEF-H } \\
\vspace{3ex}

{\Large Amsterdam NL} \\
\vspace{5ex}

{\bf September 14, 1994} \\
\vspace{15ex}

{\small {\bf Abstract}}

\end{center}

\nit
{\small
New kinds of supersymmetry arise in supersymmetric $\sg$-models describing
the motion of spinning point-particles in classical backgrounds, for example
black-holes, or the dynamics of monopoles. Their geometric origin is the
existence of Killing-Yano tensors. The relation between these concepts is
explained and examples are given.}

\np

\pagestyle{plain}
\pagenumbering{arabic}

\section{Spinning particles and supersymmetric $\sg$-models \label{S.1}}

At low energies, when back reaction effects may be neglected, the dynamics of
spinning point-particles in a $D$-dimensional curved space-time is described by
the one-dimensional supersymmetric $\sg$-model \ct{BM}-\ct{JW} with Lagrangian

\be
L\, =\, \frac{1}{2} g_{\mu\nu}(x) \dot{x}^{\mu} \dot{x}^{\nu}\, +\,
        \frac{i}{2} \eta_{ab} \ps^a \frac{D\ps^b}{D\tau}.
\label{1.1}
\ee

\nit
The position co-ordinates $x^{\mu}$ are Grassmann-even, whilst the spin
co-ordinates $\ps^a$ are Grassmann-odd and transform as a tangent-space
Lorentz vector. The standard supersymmetry variations, under which $L$
transforms into a total derivative, are

\be
\del x^{\mu}\, =\, -i\eps\ps^{\mu}, \hspace{4em}
\del \ps^{\mu}\, =\, \dot{x}^{\mu} \eps,
\label{1.2}
\ee

\nit
where we have used the notation $\ps^{\mu} = e^{\mu}_{\:\:a} \ps^a$ with
$e^{\mu}_{\:\:a}$ the inverse vielbein. The anti-symmetric spin tensor
containing the electric and magnetic dipole moments \ct{JW2} is

\be
S^{ab}\, =\, -i \ps^a \ps^b.
\label{1.2.1}
\ee

\nit
In a covariant canonical formulation \ct{GRvH,JW3} the dynamics follows from
the Hamiltonian

\be
H\, =\, \frac{1}{2}\, g^{\mu\nu} \Pi_{\mu} \Pi_{\nu}.
\label{1.3}
\ee

\nit
Here $\Pi_{\mu}$ is the covariant momentum, which can be expressed in terms
of the canonical momentum $p_{\mu}$ and the spin-connection $\og_{\mu ab}$
as

\be
\Pi_{\mu}\, =\, p_{\mu}\, -\, \frac{1}{2}\, \og_{\mu ab} S^{ab}\,
            =\, g_{\mu\nu} \dot{x}^{\nu}.
\label{1.4}
\ee

\nit
In this formulation the time evolution of any dynamical quantity
$F(x,\Pi,\ps)$ can be computed in terms of a Poisson-Dirac bracket

\be
\dot{F}\, =\, \left\{ F, H \right\},
\label{1.4.1}
\ee

\nit
with the non-vanishing elementary Poisson-Dirac brackets given by

\be
\ba{lllllll}
\left\{ x^{\mu}, \Pi_{\nu} \right\} & = & \del^{\mu}_{\nu}, & \hspace{3em} &
\left\{ \Pi_{\mu}, \Pi_{\nu} \right\} & = & R_{\mu\nu} \equiv
                                            \frac{1}{2}\, S^{ab} R_{ab\mu\nu},
\\
 & & & & & & \\
\left\{ \ps^a, \ps^b \right\} & = & \eta^{ab}, &  &
\left\{ \Pi_{\mu}, \ps^a \right\} & = & - i\og_{\mu \:\: b}^{\:\:a} \ps^b. \\
\ea
\label{1.5}
\ee

\nit
Note that $R_{\mu\nu}$ are the components of the spin-valued Riemann tensor. It
follows, that for general scalar phase-space functions $F$ and $G$ the brackets
read

\be
\left\{ F, G \right\}\, =\, \cD_{\mu}F\dd{G}{\Pi_{\mu}}\, -\, \dd{F}{\Pi_{\mu}}
\cD_{\mu} G\, +\, R_{\mu\nu} \dd{F}{\Pi_{\mu}} \dd{G}{\Pi_{\nu}}\, +\,
i (-1)^{a_F}\, \dd{F}{\ps^a} \dd{G}{\ps_a},
\label{1.6}
\ee

\nit
with the covariant derivatives defined by

\be
\cD_{\mu} F\, =\, \pl_{\mu} F\, +\, \Gam_{\mu\nu}^{\:\:\:\:\:\lb} \Pi_{\lb}
                  \dd{F}{\Pi_{\nu}}\, +\, \og_{\mu ab} \ps^b \dd{F}{\ps_a}.
\label{1.7}
\ee

\nit
Of course, any tensor-valued quantity can be converted into a scalar by
decomposing its tangent-space components into irreducible representations of
the
$D$-dimensional Lorentz group and saturating the symmetric index sets with
$\Pi_a = e^{\mu}_{\:\:a} \Pi_{\mu}$, and the anti-symmetric index sets with
$\ps^a$; then $F$ takes the form

\be
F(x,\Pi,\ps)\, =\, \sum_{m,n\geq 0}\, \frac{ i^{\left[ \frac{m}{2} \right]} }{
m!n!}\, \ps^{a_1} ... \ps^{a_m}\, f^{\mu_1 ... \mu_n}_{a_1 ... a_m}(x)\,
        \Pi_{\mu_1} ... \Pi_{\mu_n}.
\label{1.8}
\ee

\nit
In this way the phase-space functions become generalized differential forms
on a graded space.

As a simple application of the algebra (\ref{1.5}) we compute the brackets of
the spin tensor; we verify that it generates a realization of the Lorentz
algebra

\be
\left\{ S^{ab}, S^{cd} \right\}\, =\, \eta^{ad} S^{bc}\, +\, \eta^{bc} S^{ad}\,
                                     -\eta^{ac} S^{bd}\, -\, \eta^{bd} S^{ac}.
\label{1.9}
\ee

\nit
This confirms the interpretation of $S^{ab}$ as representing the spin of the
particle.

\section{Symmetries \label{S.4}}

In the hamiltonian formulation, symmetry transformations are generated by
the constants of motion through the Poisson-Dirac brackets. In particular, the
supersymmetry transformations (\ref{1.2}) are obtained from the conserved
supercharge

\be
Q = \Pi \cdot \ps = e^{\mu}_{\:\:a}\, \Pi_{\mu} \ps^a,
\label{4.1}
\ee

\nit
by taking the bracket

\be
\del F\, =\, i \left\{ F, Q  \right\}\, \eps.
\label{4.2}
\ee

\nit
That $Q$ is conserved and the super-transformations (\ref{4.2}) represent a
symmetry is a consequence of the bracket relations

\be
\left\{ Q, Q \right\}\, =\, - 2i H, \hspace{3em}
\left\{ Q, H \right\}\, =\, 0.
\label{4.3}
\ee

\nit
The second relation, which follows from the first by the Jacobi identity,
implies at the same time the conservation of $Q$ and the invariance of $H$
under the transformations (\ref{4.2}).

After the pattern established for the supercharge $Q$, we can now investigate
the full set of symmetries for a given space-time by solving the equation

\be
\left\{ J, H \right\}\, =\, \Pi^{\mu} \lh \cD_{\mu} J + R_{\mu\nu}\,
 \dd{J}{\Pi_{\nu}} \rh\, =\,  0,
\label{4.4}
\ee

\nit
which give all constants of motion $J(x,\Pi,\ps)$. This equation is the
generalization of the usual Killing equation to spinning space \ct{RJW1}.
However, unlike the usual case, in which the solutions of the Killing
equation are single, completely symmetric tensors, here the solutions consist
of linear combinations of symmetric tensors of different rank:

\be
J(x,\Pi,\ps)\, =\, \sum_{n \geq 0}\, \frac{1}{n!}\, J^{\mu_1 ... \mu_n}(x,\ps)
                   \Pi_{\mu_1} ... \Pi_{\mu_n},
\label{4.5}
\ee

\nit
subject to the conditions

\be
J_{\lh \mu_1 ... \mu_n ;\mu_{n+1}\rh}\, + \og_{\lh \mu_{n+1} \rd}^{\:\:\:ab}\,
       \ps_{b}\, \dd{J_{\ld \mu_1 ... \mu_n \rh}}{\ps^a}\, =\,
       - \frac{i}{2}\, \ps^a \ps^b R_{ab\:\lh\mu_{n+1}\rd}^{\:\:\:\:\:\nu}\,
       J_{\ld\mu_1 ... \mu_n \rh \nu}.
\label{4.6}
\ee

\nit
A sufficient, though not necessary, condition for a solution of the generalized
Killing equations is superinvariance of a dynamical variable:

\be
\left\{ J, Q \right\}\, =\, \ps \cdot \cD J\, +\, i \Pi \cdot \dd{J}{\ps}\,
                        =\, 0.
\label{4.7}
\ee

\nit
This equation may be considered as a kind of square root of the generalized
Killing equation. The new supersymmetries we present later satisfy this
superinvariance condition. An important exception is however the supercharge
$Q$ itself; according to (\ref{4.3}) its bracket gives the hamiltonian, which
generates proper-time translations.

The solutions of the generalized Killing equation (\ref{4.4}) are of two
distinct types: {\em generic} ones, which exist for any spinning particle
model (\ref{1.1}), and {\em non-generic} ones, which depend on the specific
background space-time considered. To the first class belong supersymmetry
and proper-time translations, generated by the supercharge and hamiltonian,
respectively. In addition there also is a `chiral' symmetry, generated by the
conserved charge

\be
\Gam_*\, =\, - \frac{i^{\left[ \frac{d}{2} \right]}}{d!}\,
             \ve_{a_1 ... a_d}\, \ps^{a_1} ... \ps^{a_d},
\label{4.8}
\ee

\nit
and a dual supersymmetry generated by

\be
Q^* \, =\, i\, \left\{ Q, \Gam_* \right\}\, =\,
           - \frac{i^{\left[ \frac{d}{2} \right]}}{(d-1)!}\, e^{\mu a_1}\,
           \Pi_{\mu}\, \ve_{a_1 ... a_d}\, \ps^{a_2} ... \ps^{a_d}.
\label{4.9}
\ee

\nit
Note that $Q^*$ is Grassmann odd in even-dimensional space-times and Grassmann
even in odd-dimensional space-times. In the special case $d=2$ dual
supersymmetry is a real supersymmetry, in the sense that the bracket of $Q^*$
with itself closes on the hamiltonian. For all $d > 2$ this bracket vanishes
identically.

\section{New supersymmetries \label{S.5}}

The existence of non-generic symmetries depends by definition on the
background space-time considered. We now ask, what are the necessary
conditions for the existence of new supersymmetries such that

\be
\del x^{\mu}\, =\, - i \eps\, f^{\mu}_{\:\:a}(x) \ps^a,
\label{5.1}
\ee

\nit
with $f^{\mu}_{\:\:a}$ some vector not equal to the vierbein $e^{\mu}_{\:\:a}$.
It is straightforward to establish that the solution to this problem is the
existence of a constant of motion

\be
Q_f\, =\, f^{\mu}_{\:\:a}\, \Pi_{\mu} \ps^a\, +\, \frac{i}{3!}\, c_{abc}\,
          \ps^a \ps^b \ps^c,
\label{5.2}
\ee

\nit
with the tensorial quantities $f^{\mu}_{\:\:a}$ and $c_{abc}$ subject to

\be
\ba{c}
D_{\mu} f^{\:\:a}_{\nu}\, +\, D_{\nu} f^{\:\:a}_{\mu}\, =\, 0, \\
  \\
D_{\mu} c_{abc}\, =\, -\, R_{\mu\nu ab} f^{\nu}_{\:\:c}\, -\, R_{\mu\nu bc}
                  f^{\nu}_{\:\:a}\, -\, R_{\mu\nu ca} f^{\nu}_{\:\:b}.
\ea
\label{5.3}
\ee

\nit
These conditions express the contents of the generalized Killing equation
(\ref{4.4}) for $Q_f$. The existence of new supersymmetries of this kind then
implies automatically the existence of new Grassmann-even constants of motion
$Z$, obtained by taking the brackets of the $Q_f$ with themselves. Let us
consider the case of $r$ new supersymmetries with generators $Q_i$, $i = 1,
..., r$, each of the general form (\ref{5.2}); then the bracket of two
supercharges reads

\be
\left\{ Q_i, Q_j \right\}\, =\,- 2i Z_{ij},
\label{5.4}
\ee

\nit
with $Z_{ij}$ a Grassmann-even quadratic expression in the momenta $\Pi_{\mu}$:

\be
Z_{ij}\, =\, \frac{1}{2}\, K^{\mu\nu}_{ij}\, \Pi_{\mu} \Pi_{\nu}\, +\,
             I^{\mu}_{ij} \Pi_{\mu}\, +\, G_{ij}.
\label{5.5}
\ee

\nit
The explicit expressions for co-efficients of the three terms are

\be
\ba{lll}
K^{\mu\nu}_{ij} & = & K^{\nu\mu}_{ij}\, =\, \frac{1}{2}\,\lh f^{\mu}_{i\:a}
                      f^{\nu a}_j + f^{\mu}_{j\:a} f^{\nu a}_i \rh, \\
  & & \\
I^{\mu}_{ij} & = & \frac{i}{2}\, \ps^a \ps^b \lh f^{\nu}_{i\:b} D_{\nu}
                   f^{\mu}_{j\:a}\, +\, f^{\nu}_{j\:b} D_{\nu} f^{\mu}_{i\:a}\,
                   +\, \frac{1}{2} f^{\mu c}_i c_{j\:abc}\, +\, \frac{1}{2}
                   f^{\mu c}_j c_{i\:abc}  \rh, \\
  & & \\
G_{ij} & = & - \frac{1}{4}\, \ps^a \ps^b \ps^c \ps^d \lh R_{\mu\nu ab}
             f^{\mu}_{i\:c} f^{\nu}_{j\:d}\, +\,
             \frac{1}{2}\, c_{i\:ab}^{\:\:\:\:\:e} c_{j\:cde} \rh.
\ea
\label{5.6}
\ee

\nit
Since $Q_i$ are conserved, the Jacobi identities for the brackets (\ref{5.4})
imply the conservation of the bosonic charges $Z_{ij}$:

\be
\frac{dZ_{ij}}{d\tau}\, =\, \left\{ Z_{ij}, H \right\}\, =\, 0.
\label{5.8}
\ee

\nit
Since this is a linear relation, we may regard the $Z_{ij}$ as the components
of
a matrix $Z$ of constants of motion, each of which is a solution of the
generalized Killing equations (\ref{4.4}). The matrix-valued co-efficients
of the various terms in the momentum expansion of $Z$ then satisfy the linear
relations (\ref{4.6}):

\be
\ba{lll}
K_{ \lh \mu \nu ; \lb \rh } & = & 0, \\
  &  &  \\
D_{\lh \mu \rd} I_{\ld \nu \rh ab} & = & R_{ab\lb \lh \mu \rd}
                K_{\ld \nu \rh}^{\:\:\lb}, \\
  &  &  \\
D_{\mu} G_{abcd} & = & R_{\lb\mu\left[ab \rd}\, I^{\lb}_{\:\ld cd \right]}.
\ea
\label{5.9}
\ee

\nit
In these equations parentheses denote symmetrization and the square brackets in
the last expression denote anti-symmetrization over the (latin) indices
enclosed.

\section{Killing-Yano tensors \label{S.6}}

According to eqs.(\ref{5.2}), a generalized supersymmetry $Q_f$ exists if and
only if we can find a pair of tensors $(f^{\mu}_{\:\:a}, c_{abc})$ satisfying
the differential relations (\ref{5.3}). We now show that if $Q_f$ is
superinvariant:

\be
\left\{ Q, Q_f \right\}\, =\, 0,
\label{6.1}
\ee

\nit
then the conditions for the presence of a new supersymmetry in the $\sg$-model
(\ref{1.1}) reduce to the existence of an anti-symmetric tensor $f_{\mu\nu}$
such that

\be
f_{\mu\nu ;\lb} + f_{\lb\nu ;\mu}\, =\, 0.
\label{6.3}
\ee

\nit
This equation itself is equivalent to the first eq.(\ref{5.3}), rewritten in
terms of

\be
f_{\mu\nu}\, =\, f_{\mu}^{\:\:a} e_{\nu a}.
\label{6.3.1}
\ee

\nit
The anti-symmetry of the tensor $f_{\mu\nu}$ is a consequence of condition
(\ref{6.1}). An anti-symmetric tensor of this type is called a Killing-Yano
tensor.

Having a Killing-Yano tensor guarantees the existence of an anti-symmetric
3-index Lorentz tensor $c_{abc}$ satisfying the second equation (\ref{5.3}).
Indeed, we observe that the anti-symmetry of $f_{\mu\nu}$ in
combination with the constraint (\ref{6.3}) implies that its covariant
derivative is completely anti-symmetric:

\be
\ba{lll}
H_{\mu\nu\lb} & = & \dsp{ \frac{1}{3}\, \lh f_{\mu\nu ;\lb} + f_{\nu\lb ;\mu}
                            + f_{\lb\mu ;\nu} \rh }\\
  &  &  \\
              & = & f_{\mu\nu;\lb}.
\ea
\label{6.4}
\ee

\nit
Therefore the field strength of $f_{\mu\nu}$ is a pure gradient. Further
differentiation of $H_{\mu\nu\lb}$ and use of the Ricci identity then leads to
the result

\be
H_{\mu\nu\lb ;\kg}\, =\, \frac{1}{2}\, \lh R_{\mu\nu\kg}^{\:\:\:\:\:\:\:\sg}
   f_{\sg\lb} + R_{\nu\lb\kg}^{\:\:\:\:\:\:\:\sg} f_{\sg\mu} +
   R_{\lb\mu\kg}^{\:\:\:\:\:\:\:\sg} f_{\sg\nu} \rh.
\label{6.5}
\ee

\nit
Comparing with the second Killing equation (\ref{5.3}) we conclude, that
it is solved by taking

\be
c_{abc}\, =\, - 2 H_{abc}
\label{6.6}
\ee

\nit
for the local Lorentz 3-form corresponding to the field strength tensor. Thus,
given a Killing-Yano tensor $f_{\mu\nu}$, an anti-symmetric 3rd rank tensor
of the desired type always exists.

We conclude, that the existence of a Killing-Yano tensor is both a necessary
and a sufficient condition for the existence of a new supersymmetry of the type
(\ref{5.2}) obeying the superinvariance condition (\ref{6.1}).

\section{N-extended supersymmetry \label{S.8}}

A special case of a new supersymmetry $Q_f$ satisfying eq.(\ref{6.1}) is
that of an additional conventional supersymmetry, for which the bracket closes
on the Hamiltonian. Consider the simplest case, in which there is one
independent new supersymmetry, generated by a charge $\tQ$:

\be
\left\{ Q, \tQ \right\}\, =\, 0, \hspace{3em}
\left\{ \tQ, \tQ \right\}\, =\, -2i H.
\label{8.1}
\ee

\nit
Since the bosonic constant of motion $Z$ now coincides with $H$, we have

\be
K^{\mu\nu}\, =\, g^{\mu\nu},
\label{8.2}
\ee

\nit
whilst the other components of $Z$ vanish. As a result we have

\be
f^{\mu}_{\:\:a} f_{\nu}^{\:\:a}\, =\, \del_{\nu}^{\mu}.
\label{8.3}
\ee

\nit
The anti-commutativity of the two independend supercharges requires the
anti-symmetry of $f_{\mu\nu}$ as before. Therefore we can rewrite
eq.(\ref{8.3})
in the form

\be
f^{\mu}_{\:\:\lb} f^{\lb}_{\:\:\nu}\, =\, - \del_{\nu}^{\mu}.
\label{8.5}
\ee

\nit
Moreover, since the covariant hamiltonian contains no explicit $\ps^4$-terms,
there is no $\ps^3$-term in $\tQ$:

\be
c_{abc} = 0, \hspace{2em} \Rightarrow \hspace{2em}
f^{\mu b} f_{\nu\:\: ;\mu}^{\:\:a}\, =\, f^{\mu b} f_{\nu\:\: ;\mu}^{\:\:a}.
\label{8.6}
\ee

\nit
We conclude, that the existence of a second conventional supersymmetry requires
a complex structure $f_{\mu\nu}$, and this restricts the manifolds on which the
models are defined to be of K\"{a}hler type. Hence in this case our general
conditions reduce to the well-known standard requirements for $N=2$
supersymmetry \ct{AF}.

For higher $N$-extended supersymmetry these arguments are easily generalized.
In particular, it requires the existence of $N$ independend and mutually
anti-commuting complex structures $f^{\mu}_{i\, \nu}$ $(i=1,...,N)$:

\be
f^{\mu}_{i\, \lb} f^{\lb}_{j\, \nu}\, +\, f^{\mu}_{j\, \lb} f^{\lb}_{i\, \nu}\,
  =\, -2 \del_{ij} \del^{\mu}_{\nu}.
\label{8.6.1}
\ee

\nit
This is an $N$-dimensional pseudo-Clifford algebra. For example, for $N = 4$ it
becomes the quaternion algebra. Therefore theories with $N=4$ supersymmetry can
be realized only on target manifolds of hyper-K\"{a}hler type. In this way we
reobtain the well-known connection between supersymmetry and the division
algebras.

We can now also understand why an $N = 2$ supersymmetry is always
present in 2-dimensional target space-time. For $D = 2$ the invariant
anti-symmetric Lorentz tensor $\eps_{ab}$ provides a {\em generic} Killing-Yano
tensor

\be
f^{\mu}_{\:\:a}\, =\, e^{\mu b} \eps_{ba} , \hspace{3em}
H_{\mu\nu\lb}\, =\, D_{\lb}\, \lh e^{-1} \eps_{\mu\nu} \rh\, =\, 0.
\label{8.7}
\ee

\nit
Since the $\eps$-symbol is covariantly constant,  the corresponding 3-index
tensor $c_{abc}$ vanishes. The supercharge constructed with this special
Killing-Yano tensor in $D=2$ is then precisely the dual supercharge $Q^*$.

\section{Examples \label{S.9}}

Except for the application to $N$-extended supersymmetry, there are also
interesting examples of genuine generalized supersymmetries, which do not close
on the Hamiltonian but on charges constructed out of other symmetric
Stackel-Killing tensors. Here we give two examples, one pertaining to
black-hole geometry, and one relevant to the theory of monopoles.

The first example is provided by the D=4 Kerr-Newman black holes. These
are solutions of the combined Einstein-Maxwell equations; the line-element
reads

\be
\ba{lll}
ds^2 & = &\dsp{ - \frac{\Del}{\rg^2}\, \lh dt - a \sin^2 \th d \fg \rh^2\, +\,
             \frac{\sin^2 \th}{\rg^2}\, \lh (r^2 + a^2)d\fg - a dt \rh^2 }\\
  &  & \\
  &  & \dsp{ +\, \frac{\rg^2}{\Del}\, dr^2\, +\, \rg^2 d\th^2. }
\ea
\label{9.1}
\ee

\nit
We have used the abbreviations

\be
\rg^2\, =\, r^2\, +\, a^2\, \cos^2 \th, \hspace{3em}
\Del\, =\, r^2\, +\, a^2\, - 2 Mr\, +\, e^2\, >\, 0.
\label{9.2}
\ee

\nit
In case of non-vanishing charge $e$ the corresponding electro-magnetic field
is described by the Maxwell 2-form

\be
\ba{lll}
F & = & \dsp{ \frac{e}{\rg^4}\, \lh r^2 - a^2 \cos^2 \th \rh\, dr\, \wig\,
    \lh dt - a \sin^2 \th d\fg \rh }\\
  & & \\
  & & \dsp{ +\, 2\, \frac{ear\cos \th \sin \th}{\rg^4}\, d\th\, \wig\,
            \lh (r^2 + a^2) d\fg - a dt \rh. }
\ea
\label{9.3}
\ee

\nit
For this geometry one can find a Killing-Yano tensor $f_{\mu}^{\:\:a}$
\ct{P,F} with components

\be
\ba{lll}
f_{\mu}^{\:\:0}\, d x^{\mu} & = & \dsp{ \frac{\rg}{\Del}\, \cos \th\, dr, }\\
  &  & \\
f_{\mu}^{\:\:1}\, d x^{\mu} & = & \dsp{ - \frac{\sqrt{\Del}}{\rg}\, \cos \th\,
                                  \lh dt - a \sin^2 \th\, d\fg \rh, }\\
  &  & \\
f_{\mu}^{\:\:2}\, d x^{\mu} & = & \dsp{ - \frac{r \sin \th}{a \rg}\,
                                  \lh (r^2 + a^2) d\fg - a dt \rh, }\\
  &  & \\
f_{\mu}^{\:\:3}\, d x^{\mu} & = & \dsp{ - \frac{r \rg}{a}\, d\th.}
\ea
\label{9.4}
\ee

\nit
The corresponding components of the Lorentz 3-form $c_{abc}$ obtained from the
field strength are:

\be
\ba{lll}
c_{012} & = & \dsp{ \frac{2\sin\th}{\rg}, }\\
 & & \\
c_{123} & = & \dsp{ -\frac{2 \sqrt{\Del}}{a \rg}, }\\
 & & \\
c_{013} & = & c_{023}\, =\, 0.
\ea
\label{9.5}
\ee

\nit
The bosonic constant of motion $Z$ then contains the symmetric Stackel-Killing
tensor \ct{BC}

\be
\ba{lll}
\frac{1}{2}\, K_{\mu\nu} \dot{x}^{\mu} \dot{x}^{\nu} & = & \dsp{
   \frac{\Del \cos^2 \th}{\rg^2}\, \lh \dot{t} - a \sin^2 \th
   \dot{\fg} \rh^2\, +\, \frac{r^2 \sin^2 \th}{\rg^2 a^2}\,
   \lh (r^2 + a^2) \dot{\fg} - a\dot{t} \rh^2 }\\
  &  & \\
  &  & \dsp{ -\, \frac{\rg^2 \cos^2 \th}{\Del}\, \dot{r}^2\, +\,
             \frac{\rg^2}{r^2}{a^2}\, \dot{\th}^2,}
\ea
\label{9.6}
\ee

\nit
which is the square of the Killing-Yano tensor. \nl

\nit
Our second example concerns the dynamics of magnetic monopoles. As shown in
\ct{M,GM} the effective action for the scattering of slowly-moving monopoles is
given by the self-dual Taub-NUT solution of the D=4 Euclidean Einstein
equations
(with negative mass). The relevant Lagrangian for large distances\footnote{In
dimensionless reduced co-ordinates: $r \gg 2$.} reads

\be
L\, =\, \pi \lh \frac{ds}{d\tau} \rh^2,
\label{9.8}
\ee

\nit
where the line element in spherical co-ordinates is given by

\be
ds^2\, =\, \lh 1 - \frac{2}{r} \rh \left[ dr^2 + r^2 \lh d\th^2 + \sin^2 \th
           d \fg^2 \rh \right]\, +\, \frac{1}{\lh 1 - \frac{2}{r} \rh}\,
           \left[ d\ps + \cos \th d\fg \right]^2.
\label{9.9}
\ee

\nit
The supersymmetric extension of this Lagrangian was investigated in \ct{V1,DB}.

The metric (\ref{9.9}) admits four Killing vectors, which transform as a scalar
and a vector under rotations; they represent the relative charge $q$ and the
total angular momentum $\vec{J}$ (which includes a contribution from the
relative electric charge).

As observed in \ct{GR}, the Taub-NUT geometry also possesses four
Killing-Yano tensors. Three of these are rather special: they are covariantly
constant, mutually anti-commuting and square to minus unity:

\be
f_i f_j + f_j f_i = -2 \del_{ij}, \hspace{3em} D_{\mu} f^{\nu}_{i\:\lb}\, =\,
0.
\label{9.9.1}
\ee

\nit
Thus they are complex structures realizing the quaternion algebra. Indeed, the
Taub-NUT manifold defined by (\ref{9.9}) is hyper-K\"{a}hler and as a
consequence the corresponding supersymmetric $\sg$-model has an $N = 4$
supersymmetry. We also observe, that these three complex structures transform
as a vector under rotations generated by $\vec{J}$.

Since the Killing-Yano tensors $f_i$ are covariantly constant, their field
strengths vanish and so do the corresponding 3-index tensors $c_{i\;abc}$.
Therefore we find three vector-like supercharges of the form

\be
Q_i\, =\, f^{\mu}_{i\:a} \Pi_{\mu} \ps^a.
\label{9.10}
\ee

\nit
Denoting the original supersymmetry by $Q_0$, the complete set of brackets
of the $Q_A$, $A = 0,...,3$ realizes the $N = 4$ supersymmetry algebra:

\be
\left\{ Q_A, Q_B \right\}\, =\, -2i \del_{AB}\, H.
\label{9.11}
\ee

\nit
There are no fermionic components on the right-hand side. This is consistent
because of the properties (\ref{9.9.1}), which imply by the Ricci identity
and the self-duality of the Taub-NUT geometry that

\be
\eps^{abcd}\, R_{cd \mu\nu}\, f^{\mu}_{i\:a}\, f^{\nu}_{j\:b}\,  =\, 0.
\label{9.13}
\ee

\nit
Eqs.(\ref{9.9.1},\ref{9.13}) are sufficient to show the vanishing of the
components $I^{\mu}_{AB}$ and $G_{AB}$ on the right-hand side of
the bracket (\ref{9.11})

In addition to the vector-like Killing-Yano tensors there also is a scalar one,
called $Y_{\mu \nu}$, which has a non-vanishing field strength and,
consequently, 3-index tensor $c_{abc}$. The supercharge $Q_Y$ constructed out
of $Y$ is a scalar under rotations generated by the total angular momentum. It
now turns out \ct{GR} that the only bosonic constants of motion which contain
new dynamical information are those obtained from the brackets

\be
\left\{ Q_i, Q_Y \right\}\, =\, -2i Z_i.
\label{9.14}
\ee

\nit
The Stackel-Killing tensors $K^{\mu\nu}_i$ appearing on the right-hand side are
those found in \ct{DR}, forming a Runge-Lenz-like vector $\vec{K}$. We note in
passing, that although the $Z_i$ have a non-trivial contribution from spin
\ct{V2}, the scalar part $G_i$ still vanishes due to the identity (\ref{9.13})
and the vanishing of $c_{i\:abc}$.

\section{Conclusion \label{S.10}}

{}From the examples given it is clear, that the concepts of Killing-Yano
tensors
and the generalized supersymmetries play an important role in clarifying the
motion of fermions in the presence of black holes and monopoles. In this paper
we have concentrated on the pseudo-classical aspects, but they have their
direct
translation in the quantum theory, where the supercharges are represented by
Dirac operators

\be
Q_f\, \rightarrow\, -i \gam^a\, \lh f^{\mu}_{\:\:a} D_{\mu}\, -\,
                    \frac{1}{3!}\, c_{abc} \sg^{bc} \rh.
\label{10.1}
\ee

\nit
The corresponding Laplacians which are appropriate extensions of

\be
\Box_K\, =\, D_{\mu} K^{\mu\nu} D_{\nu},
\label{10.2}
\ee

\nit
represent the constants of motion $Z$. Some of these operators and their
quantum mechanical interpretations have been studied in the literature
\ct{BC,F,P,M,GM}. The pseudo-classical treatment here provides an alternative
road to quantization through the path-integral formulation. In both cases,
the correspondence principle leads to equivalent algebraic structures.
\vspace{5ex}

\nit
{\bf Acknowledgement}\nl

\nit
The research described in this paper is supported in part by the Human
Capital and Mobility Program through the network on {\em Constrained
Dynamical Systems}.

\end{document}